\documentclass[conference]{IEEEtran}
\IEEEoverridecommandlockouts

\usepackage[T1]{fontenc}
\usepackage[utf8]{inputenc}

\usepackage{graphicx}
\usepackage{amsmath}
\usepackage{cite}
\usepackage{url}
\usepackage{comment}
\usepackage{booktabs}
\usepackage{placeins}
\usepackage{float}
\usepackage{tikz}
\usetikzlibrary{arrows.meta,positioning}
\usepackage{tabularx}
\usepackage{array}
\usepackage{adjustbox}
\usepackage{rotating}
\usepackage{multicol}
\newcolumntype{Y}{>{\raggedright\arraybackslash}X}
\graphicspath{{figs/}{yahboom/data/}}
\usepackage[hidelinks]{hyperref}

\begin{document}

\title{Experimental Evaluation of Security Attacks on Self-Driving Car Platforms}

\author{\IEEEauthorblockN{Viet K. Nguyen, Nathan Lee, Mohammad Husain}
\IEEEauthorblockA{Cal Poly Pomona, Pomona CA 91768, USA\\
Email: vietknguyen@cpp.edu, nathanlee@cpp.edu, mihusain@cpp.edu}}

\maketitle

\begin{abstract}
Deep learning-based perception pipelines in autonomous ground vehicles are vulnerable to both adversarial manipulation and network-layer disruption. We present a systematic, on-hardware experimental evaluation of five attack classes: FGSM, PGD, man-in-the-middle (MitM), denial-of-service (DoS), and phantom attacks on low-cost autonomous vehicle platforms (JetRacer and Yahboom). Using a standardized 13-second experimental protocol and comprehensive automated logging, we systematically characterize three dimensions of attack behavior: (i) control deviation, (ii) computational cost, and (iruntime responsiveness. Our analysis reveals that distinct attack classes produce consistent and separable ``fingerprints'' across these dimensions: perception attacks (MitM output manipulation and phantom projection) generate high steering deviation signatures with nominal computational overhead, PGD produces combined steering perturbation and computational load signatures across multiple dimensions, and DoS exhibits frame rate and latency degradation signatures with minimal control-plane perturbation. We demonstrate that our fingerprinting framework generalizes across both digital attacks (adversarial perturbations, network manipulation) and environmental attacks (projected false features), providing a foundation for attack-aware monitoring systems and targeted, signature-based defense mechanisms.

\end{abstract}


\section{Introduction}
Modern autonomous vehicles rely on deep neural networks for perception and control, exposing new attack surfaces that span input-space perturbations, software pipeline manipulation, and network-layer disruption. While existing research has extensively analyzed individual attack types, there remains a critical gap in understanding how different attack classes manifest in system behavior on real autonomous vehicle hardware.

We focus on road-following tasks that process camera images to generate steering-related control signals. This task provides an ideal testbed because it relies on convolutional neural networks (CNNs) that represent the state-of-the-art for image understanding while maintaining computational efficiency suitable for embedded GPUs. In our experimental setup, a ResNet-18-based model predicts steering-related outputs from road images, creating a realistic deployment scenario that matches both the data modality and computational constraints of commercial autonomous vehicle systems.

\textbf{Contributions.} This paper makes three primary contributions to the field of autonomous vehicle security:

\begin{itemize}
    \item \textbf{Attack Fingerprinting Framework:} A systematic three-dimensional characterization of attack impact encompassing control deviation, computational cost, and system responsiveness, evaluated on live autonomous vehicle hardware.

    \item \textbf{Rigorous On-Hardware Evaluation:} A standardized 13-second experimental protocol with comprehensive automated logging, enabling consistent and reproducible evaluation across multiple attack classes with statistical rigor.

    \item \textbf{Defense-Relevant Insights:} Empirical evidence demonstrating that attack classes exhibit separable behavioral signatures that can inform the design of targeted detection and response mechanisms.
\end{itemize}

\textbf{Research Questions.} Our investigation addresses three fundamental questions about attack behavior in autonomous vehicle systems:

\begin{itemize}
    \item \textbf{RQ1:} What are the characteristic behavioral signatures of perception attacks (adversarial perturbations, sensor corruption, environmental manipulation) and network/system attacks (control hijacking, timing disruption) across control, computational, and responsiveness dimensions on physical autonomous vehicle platforms?
    \item \textbf{RQ2:} What computational and runtime behavioral patterns (e.g., FPS characteristics, latency profiles) do these attacks produce under real-world deployment conditions?
    \item \textbf{RQ3:} Do different attack classes produce consistent, discriminative behavioral "fingerprints" across these measurement dimensions that could enable practical detection and classification?
\end{itemize}

\section{System Model}

Our evaluation targets camera-based autonomous road-following systems that process visual input to generate steering commands for vehicle control. We define the system architecture, data flow, and operational constraints that characterize our experimental platforms.

\subsection{System Architecture}
The autonomous vehicle system consists of three primary components operating in a closed-loop configuration:

\textbf{Perception Module:} A convolutional neural network (CNN) processes camera images to extract road geometry information. The network employs a ResNet-18 backbone modified with a 2-neuron regression head that predicts target coordinates $(x,y)$ representing the desired road apex position within the image frame.

\textbf{Control Module:} A proportional controller converts the perception output into steering commands. The $x$-coordinate is mapped to rotation angles through a linear transformation with configurable bounds, while the $y$-coordinate influences speed control.

\textbf{Actuation Module:} Motor controllers translate the steering and speed commands into physical wheel movements. The module enforces safety bounds and implements rate limiting to prevent aggressive actuation changes.

\subsection{Data Flow Pipeline}
The system processes visual information through a well-defined pipeline:

\begin{enumerate}
    \item \textbf{Image Acquisition:} Camera captures $640 \times 480$ RGB frames at 30 FPS
    \item \textbf{Region of Interest (ROI) Extraction:} Bottom half of frame is selected to focus on road surface
    \item \textbf{Preprocessing:} ROI is resized to $224 \times 224$ and normalized using ImageNet statistics
    \item \textbf{Neural Inference:} CNN processes preprocessed tensor to generate coordinate predictions
    \item \textbf{Postprocessing:} Predictions are clipped to valid ranges and converted to control signals
    \item \textbf{Motor Control:} Commands are transmitted to wheel controllers with 50ms update intervals
\end{enumerate}

\subsection{Operational Constraints}
The experimental platforms operate under several constraints that reflect real-world deployment conditions:

\textbf{Computational Constraints:} Embedded GPU processing with limited memory bandwidth and thermal throttling considerations. Inference must complete within 33ms to maintain real-time operation.

\textbf{Timing Constraints:} End-to-end latency from image capture to motor actuation must not exceed 100ms to maintain stable control. Frame processing must maintain consistent 30 FPS throughput.

\textbf{Physical Constraints:} Maximum steering angles, acceleration limits, and safety boundaries enforced by the motor control system to prevent unsafe vehicle behavior.

\subsection{Assumptions and Scope}
We make several key assumptions about the system under evaluation:

\begin{itemize}
    \item Single-camera operation without multi-sensor fusion
    \item Structured road environments with clear lane markings
    \item Moderate vehicle speeds (1-2 m/s) ensuring stable control dynamics
    \item Controlled lighting conditions suitable for computer vision processing
    \item No adversarial training or specialized defenses implemented in baseline system
\end{itemize}

These assumptions define the scope of our evaluation and reflect typical deployment scenarios for research-grade autonomous vehicle platforms. The system model provides the foundation for understanding attack insertion points and potential impact vectors explored in subsequent sections.

\section{Background and Literature Review}
\subsection{Attack Surveys and Taxonomies}
Recent comprehensive surveys of connected and autonomous vehicle (CAV) security catalog threats across multiple dimensions: attacker knowledge (white/gray/black box), attack objectives (targeted vs. untargeted), and attack surfaces spanning sensors, perception models, control channels, and networks~\cite{kim2021survey,sun2022survey,10.1145/3691625}. These taxonomies emphasize operational manifestations such as added latency, frame drops, and biased control messages that critically impact closed-loop stability.

\subsection{Digital Adversarial Attacks for Driving}
Adversarial examples exploit model sensitivity to small perturbations in pixel space. Canonical methods include FGSM~\cite{goodfellow2015explainingharnessingadversarialexamples} and iterative PGD along with adversarial training~\cite{madry2019deeplearningmodelsresistant}. In driving contexts, Sato et al.\ demonstrated that adversarial road-surface patches can cause lane departure in automated lane centering systems within driver reaction time~\cite{sato2021dirty}. However, Wang et al.\ showed that even attacks achieving high component-level success rates (e.g., $>$70\% misdetection) can fail to produce system-level effects when evaluated within a full AD pipeline including tracking, planning, and control~\cite{wang2023sysadv}, highlighting the critical need for on-hardware, closed-loop evaluation, a central motivation of our work.

\subsection{Physical-World Transfer to Cameras}
Physical realizations demonstrate transfer from printed patterns to camera inputs: posters/patches and signage manipulations can reliably induce misclassification or steering errors under varying viewpoints and lighting~\cite{kurakin2017adversarialphysical,eykholt2018robust,brown2017adversarialpatch,sitawarin2018dartsdeceivingautonomouscars}. These results highlight risks beyond purely digital pipelines and motivate evaluations that consider sensor capture conditions.

\subsection{Network and Pipeline Attacks}
In connected autonomous vehicle stacks, man-in-the-middle, false-data injection/replay, and denial-of-service can degrade perception–control performance by biasing data streams or introducing timing jitter and frame drops~\cite{kim2021survey,sun2022survey}. Practical pipelines are vulnerable at multiple insertion points (camera stream interception, model I/O, controller interface), and even modest delays can destabilize closed-loop behavior.

\subsection{Defenses and Practical Constraints}
Defenses span robust training/optimization~\cite{madry2019deeplearningmodelsresistant}, input transformations, and runtime anomaly detection/monitoring~\cite{10.1145/3691625}. Embedded constraints (compute, thermals, tight latency budgets) complicate heavy defenses and can introduce timing side effects, motivating lightweight monitoring strategies that reason about runtime symptoms. See Section~\ref{sec:mitigations} for a fingerprint-aware mitigation blueprint aligned with our experimental findings.

\subsection{Summary and Gap}
While prior work has made substantial advances in understanding individual attack vectors and defense mechanisms, several critical gaps remain. Most studies focus on single attack classes or rely on simulation-based evaluations that cannot capture real-world deployment complexities. On-device, cross-attack comparisons that jointly characterize control deviation, computational cost, and responsiveness are particularly scarce.

Most importantly, there is a lack of systematic frameworks for understanding how different attack classes manifest distinct operational "fingerprints" that could inform practical monitoring and defense strategies. Our work addresses this gap by introducing a comprehensive three-dimensional fingerprinting framework that enables cross-attack comparison on real autonomous vehicle hardware, providing insights that simulation-based studies cannot capture.
 
\section{Methodology}

\subsection{Threat Model}

We consider a post-compromise threat model that evaluates the impact and damage potential of attacks after an adversary has gained initial access to an autonomous vehicle system. This approach reflects realistic security scenarios where attackers leverage initial system compromise to execute deeper attacks on critical vehicle functions. Our threat model operates under carefully defined assumptions about adversary capabilities, objectives, and constraints.

\subsubsection{Adversary Capabilities}
Following initial system compromise, the attacker has gained significant access to vehicle systems:

\begin{itemize}
    \item \textbf{White-box model access:} Complete knowledge of perception model architecture, parameters, and training procedures through system infiltration, insider access, or model extraction techniques. This enables gradient-based attacks and fine-grained perturbation optimization.

    \item \textbf{Network-level access:} Presence within the compromised vehicle's local network infrastructure with ability to intercept, modify, and inject network traffic between system components.

    \item \textbf{Software pipeline access:} Capability to manipulate data flow between sensors, perception models, and control systems through process injection, library hooking, or direct code modification while maintaining no direct physical access to vehicle hardware.

    \item \textbf{Timing control:} Ability to introduce delays, drop frames, or flood processing queues to affect system responsiveness and timing characteristics.
\end{itemize}

\subsubsection{Attack Objectives and Success Metrics}
In the post-compromise phase, the adversary seeks to achieve specific malicious goals while maintaining stealth:

\begin{itemize}
    \item \textbf{Control hijacking:} Manipulating vehicle steering and navigation to cause path deviation from intended trajectory. Success measured by lateral displacement from planned path and ability to maintain persistent control deviations.

    \item \textbf{System disruption:} Degrading computational performance and system responsiveness to create operational failures or safety hazards. Success measured by frame rate reduction, latency increase, and processing queue saturation.

    \item \textbf{Stealth operation:} Executing attacks while avoiding detection by system monitors, operators, or external observers. Success measured by the attack's ability to masquerade as normal system behavior or benign environmental factors.

    \item \textbf{Impact characterization:} Understanding the unique behavioral signatures of different attack vectors for sustained compromise and attack refinement.
\end{itemize}

\subsubsection{Adversary Constraints and Limitations}
The attacker operates under several realistic constraints that bound attack capabilities:

\begin{itemize}
    \item \textbf{Limited hardware access:} Cannot directly modify vehicle hardware components or physical sensors (e.g., cannot tamper with camera internals, GPS units, or onboard computers), though environmental manipulation visible to sensors is within scope (e.g., phantom projections)
    \item \textbf{Limited computational resources:} Attack generation must be feasible on commodity hardware within time constraints of real-time operation
    \item \textbf{Detection avoidance:} Excessive deviations or obvious system failures may trigger operator intervention
    \item \textbf{Environmental constraints:} Limited by physical laws, sensor capabilities, and vehicle dynamics
\end{itemize}

\subsubsection{Target Environment and Assumptions}
We evaluate compromised research platforms operating under realistic deployment conditions:

\textbf{Target Systems:} NVIDIA JetRacer and Yahboom ROSMaster X3 research platforms with default security configurations typical of academic and development environments. These systems lack enterprise-grade security measures, representing realistic post-breach scenarios.

\textbf{Operational Context:} Vehicles operate in structured environments at moderate speeds (1-2 m/s) with clear lane markings and controlled lighting. The road-following task provides a standardized evaluation scenario that enables consistent measurement of attack impact across different platforms.

\textbf{Defensive Posture:} Baseline systems operate without specialized security mechanisms or adversarial training. This represents worst-case scenarios for security but provides clean measurements of attack effects without defensive interference.

This threat model enables systematic evaluation of attack impacts while maintaining relevance to real-world deployment scenarios and security challenges faced by autonomous vehicle systems.

\subsection{Attack Fingerprinting Framework}

Our core methodological contribution is a systematic framework for characterizing the unique operational "fingerprints" of different attack classes on autonomous vehicle platforms. Each attack type produces distinct measurable signatures across three dimensions:

\begin{itemize}
    \item \textbf{Control Impact:} Measured by steering deviation from normal operation baseline
    \item \textbf{Performance Impact:} Quantified by processing time increases and computational resource consumption
    \item \textbf{System Responsiveness:} Assessed through frame rate degradation and system lag metrics
\end{itemize}

This multi-dimensional characterization enables the identification of attack-specific signatures that can inform targeted defense mechanisms, moving beyond simple binary detection to attack-type classification.

\subsection{Experimental Approach}

We implement attacks directly within the vehicle's software pipeline, manipulating dataflow between the camera, perception model, and motor controller on physical hardware. This on-device approach allows us to measure real-world performance impacts and system behavior under attack conditions, providing insights that simulation-based studies cannot capture.

\section{Attack Models and Insertion Points}
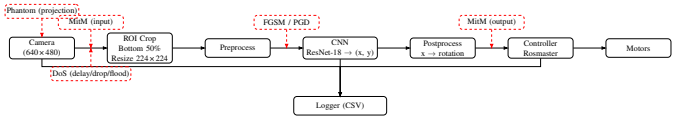
\begin{figure}[H]
    \centering
    \resizebox{\linewidth}{!}{%
    \begin{tikzpicture}[
      node distance=10mm and 14mm,
      box/.style={draw, rounded corners, align=center, inner sep=3pt, minimum width=28mm, minimum height=8mm},
      tap/.style={draw, dashed, rounded corners, align=center, inner sep=2pt, minimum width=28mm, minimum height=6mm, draw=red},
      >=Latex
    ]
    % Core pipeline
    \node[box] (cam) {Camera\\(640\,\texttimes\,480)};
    \node[box, right=of cam] (roi) {ROI Crop\\Bottom 50\%\\Resize 224\,\texttimes\,224};
    \node[box, right=of roi] (prep) {Preprocess};
    \node[box, right=of prep, minimum width=32mm] (cnn) {CNN\\ResNet-18 $\to$ (x, y)};
    \node[box, right=of cnn] (post) {Postprocess\\x $\to$ rotation};
    \node[box, right=of post] (ctrl) {Controller\\Rosmaster};
    \node[box, right=of ctrl] (mot) {Motors};

    \draw[->] (cam) -- (roi);
    \draw[->] (roi) -- (prep);
    \draw[->] (prep) -- (cnn);
    \draw[->] (cnn) -- (post);
    \draw[->] (post) -- (ctrl);
    \draw[->] (ctrl) -- (mot);

    % Logger taps
    \node[box, below=16mm of cnn, minimum width=40mm] (log) {Logger (CSV)};
    \draw[->] (cam) -- ++(0,-8mm) -| (log);
    \draw[->] (cnn) -- ++(0,-8mm) -| (log);
    \draw[->] (ctrl) -- ++(0,-8mm) -| (log);

    % Attack insertion points anchored to edges
    % Define edge midpoints for precise attachment
    \path (prep) -- (cnn) node[midway, coordinate] (e_prep_cnn) {};
    \path (cam)  -- (roi) node[midway, coordinate] (e_cam_roi)  {};
    \path (post) -- (ctrl) node[midway, coordinate] (e_post_ctrl){};

    % FGSM/PGD on Preprocess -> CNN edge
    \node[tap, above=8mm of e_prep_cnn] (fgsm) {FGSM / PGD};
    \draw[dashed,->, red] (fgsm) -- (e_prep_cnn);

    % MitM (input) on Camera -> ROI edge
    \node[tap, above=8mm of e_cam_roi] (mitmin) {MitM (input)};
    \draw[dashed,->, red] (mitmin) -- (e_cam_roi);

    % MitM (output) on Postprocess -> Controller edge
    \node[tap, above=8mm of e_post_ctrl] (mitmout) {MitM (output)};
    \draw[dashed,->, red] (mitmout) -- (e_post_ctrl);

    % DoS on Camera -> ROI edge (timing/queue disruption)
    \node[tap, below=8mm of e_cam_roi] (dos) {DoS (delay/drop/flood)};
    \draw[dashed,->, red] (dos) -- (e_cam_roi);

    % Phantom attack - before camera (physical environment)
    \node[tap, above=8mm of cam] (phantom) {Phantom (projection)};
    \draw[dashed,->, red] (phantom) -- (cam.north);
    \end{tikzpicture}%
    }

    \caption{Perception--control pipeline on the research platforms. Attacks insert either before the model (input pixel-space), between model and controller (output manipulation), along the I/O/network path (DoS/MitM), or in the physical environment before camera capture (Phantom).}
    \label{fig:pipeline_overview}
\end{figure}

We systematically characterize where each attack intervention point affects the perception-control pipeline and identify the specific system components stressed by different attack vectors. Figure~\ref{fig:pipeline_overview} illustrates the nominal dataflow and highlights the specific edges where attacks are inserted.

\subsection{FGSM (Input--Pixel)}
\textbf{Purpose:} Single-step adversarial perturbation that maximizes loss (targeted or untargeted) within an \(\ell_\infty\) bound.

\textbf{Insertion point:} Just before the perception model (after ROI crop and normalization).

\textbf{Key parameters:} \(\epsilon\) (perturbation budget); targeted vs. untargeted objective.

\textbf{Primary module impacted:} Minimal additional compute; the CNN forward path remains the main cost.

\textbf{Runtime signature:} Moderate control deviation relative to baseline; small processing-time overhead; FPS approximately unchanged.

\textbf{Logged fields (excerpt):} \texttt{attack\_type=FGSM}, \texttt{epsilon}, \texttt{adv\_x}, \texttt{adv\_y}, \texttt{processing\_time\_ms}.

\subsection{PGD (Input--Pixel, Iterative)}
\textbf{Purpose:} Iterative first-order attack that maximizes loss within an \(\ell_\infty\) ball by taking multiple gradient steps with projection.

\textbf{Insertion point:} Immediately before the perception model (after ROI crop and normalization), replacing the clean frame with an adversarial one each iteration window.

\textbf{Key parameters:} \(\epsilon\) (budget), \(\alpha\) (step size), \(N\) (number of iterations), \texttt{random\_start} (initialize within the \(\epsilon\)-ball).

\textbf{Primary module impacted:} Perception compute budget (iterative perturbation adds synchronous work within the control loop), leading to queueing and increased per-frame processing time.

\textbf{Runtime signature:} Large processing-time increases and FPS drops relative to baseline; substantial steering deviation; often high-frequency oscillations in the steering time series.

\textbf{Logged fields (excerpt):} \texttt{attack\_type=PGD}, \texttt{epsilon}, \texttt{alpha}, \texttt{num\_iter}, \texttt{random\_start}, \texttt{adv\_x}, \texttt{adv\_y}, \texttt{processing\_time\_ms}, \texttt{rapid\_steering\_event}.

\subsection{MitM -- Input (Sensor Stream)}
\textbf{Purpose:} Corrupt or replace the sensor stream before it reaches the model to induce misperception.

\textbf{Insertion point:} Along the Camera $\to$ ROI/Preprocess edge (software shim on-device, or a proxy if using a streamed feed), as indicated in Figure~\ref{fig:pipeline_overview}.

\textbf{Variants \& parameters:} Additive noise (standard deviation), blur (kernel size/sigma), or full synthetic frame replacement (source/refresh rate).

\textbf{Primary module impacted:} Perception robustness to corrupted inputs; if networked, the decode/encode path. Compute overhead is typically low.

\textbf{Runtime signature:} Visible frame corruption in logs; control deviation depends on content and intensity; processing time and FPS near baseline.

\textbf{Logged fields (excerpt):} \texttt{attack\_layer=input}, \texttt{attack\_type} in \{noise, blur, synthetic\}, \texttt{intensity}, \texttt{manipulation\_magnitude}, \texttt{frame\_saved}.

\subsection{MitM -- Output (Control Channel)}
\textbf{Purpose:} Bias or override steering/control messages after model inference to hijack actuation.

\textbf{Insertion point:} Along the Postprocess $\to$ Controller edge (or on the control channel if networked), as shown in Figure~\ref{fig:pipeline_overview}.

\textbf{Key parameters:} Target value (fixed command), bias magnitude (additive offset), randomness (random target within bounds).

\textbf{Primary module impacted:} Control interface (software I/O); minimal compute cost.

\textbf{Runtime signature:} High and consistent steering bias with negligible processing-time increase and FPS change; clear shift in time-series baseline.

\textbf{Logged fields (excerpt):} 
\texttt{attack\_layer=output\_modification, attack\_type in \{target, random\}, manipulation\_magnitude, mitm\_success, processing\_time\_ms.}

\subsection{Denial-of-Service (Timing/Throughput)}
\textbf{Purpose:} Degrade responsiveness or starve the control loop by introducing delays, dropping frames, or flooding I/O/network paths.

\textbf{Insertion point(s):} Along the Camera $\to$ ROI/Preprocess edge (capture thread sleeps or drops), at the network/AP path (traffic shaping/queue bloat), or at the application endpoint (server RX queue). See Figure~\ref{fig:pipeline_overview}.

\textbf{Subtypes \& parameters:}
\begin{itemize}
  \item \emph{Delay:} add per-frame delay (ms). Parameter: \texttt{delay\_ms} or normalized intensity.
  \item \emph{Drop:} probabilistically skip frames. Parameter: \texttt{drop\_rate} (\%).
  \item \emph{Flood:} saturate network or endpoint with excess traffic. Parameters: \texttt{flood\_rate} (pps/MBps), protocol.
\end{itemize}

\textbf{Primary module(s) overwhelmed:} Camera capture/I/O thread (scheduling), RX buffers/queues in the network stack or application, causing stale inputs to propagate to the controller.

\textbf{Runtime signature:} Large FPS drop and processing-time increase relative to baseline; visible lag in time-series; control deviation often small unless staleness is severe or compounded with other effects.

\textbf{Logged fields (excerpt):} \texttt{dos\_enabled}, \texttt{dos\_type} in \{delay, drop, flood\}, \texttt{dos\_intensity}, \texttt{dropped\_frames}, \texttt{processing\_time\_ms}, \texttt{fps\_estimate}, \texttt{performance\_degradation}.

\subsection{Phantom Attack (Environmental Perception Manipulation)}
\textbf{Purpose:} Manipulate the physical environment visible to the camera by projecting fake visual features (e.g., lane markings, obstacles) to induce misperception without directly modifying sensor hardware or digital dataflow.

\textbf{Insertion point:} Physical environment before camera capture, outside the software pipeline shown in Figure~\ref{fig:pipeline_overview}. Attack occurs in the pre-sensor physical domain rather than the digital processing pipeline.

\textbf{Implementation:} A projector positioned at the calculated 5-second position along the track projects fake lane line patterns onto the road surface. The projector activates at the attack phase boundary (t=5s) and remains active through the 8-second attack window, creating persistent false visual features that the camera captures as part of the natural scene.

\textbf{Key parameters:} Projection pattern type (lane lines, obstacles, markers), intensity/brightness, positioning relative to vehicle trajectory, timing synchronization with experimental protocol.

\textbf{Primary module impacted:} Perception robustness to environmental manipulation. Unlike digital adversarial attacks (FGSM, PGD) which perturb pixels post-capture, or sensor stream corruption (MitM Input) which modifies data in-pipeline, phantom attacks operate in the physical domain, creating false features that appear authentic to both the camera sensor and human observers.

\textbf{Runtime signature:} High steering deviation as the model naturally responds to perceived (but fake) road features; minimal processing-time overhead since no additional computation is required; FPS approximately unchanged as frame capture and processing proceed normally.

\textbf{Distinction from other perception attacks:} While FGSM/PGD create imperceptible perturbations and MitM Input introduces visible corruption, phantom attacks create \emph{plausible false features} that exploit the model's intended functionality rather than its vulnerabilities. The model operates correctly on its input; the input itself is adversarially constructed in the physical environment.

\textbf{Logged fields (excerpt):} \texttt{attack\_type=phantom}, \texttt{projection\_pattern}, \texttt{projection\_timing}, \texttt{perceived\_feature\_type}, \texttt{steering\_x}, \texttt{processing\_time\_ms}.

\subsection*{Attack Comparison}
\noindent Table~\ref{tab:attack_models_summary} contrasts attack surfaces, insertion edges, targets, parameters, and their qualitative fingerprints.

\begin{table}[H]
\centering
\caption{How attacks operate in our pipeline. Surf./Edge references Figure~\ref{fig:pipeline_overview}; Target denotes the primarily stressed component.}
\label{tab:attack_models_summary}
\scriptsize
\begin{tabular}{@{}p{1.2cm}p{1.8cm}p{1.5cm}p{2.7cm}@{}}
\toprule
\textbf{Attack} & \textbf{Surf./Edge} & \textbf{Target} & \textbf{Signature (control + compute/FPS)} \\
\midrule
FGSM (U) & Input, Prep$\to$CNN & Perception (model) & Moderate bias; small overhead; FPS normal \\
PGD (U) & Input, Prep$\to$CNN & Perception compute & Large deviation; extreme proc-time; big FPS drop; oscillation + lag \\
MitM (Input Noise) & Input, Cam$\to$ROI & Perception robustness & Low-moderate change; corruption visible; FPS baseline \\
MitM (Input Synthetic) & Input, Cam$\to$ROI & Perception robustness & Moderate change; step change in frames; FPS baseline \\
MitM (Output Target) & Output, Post$\to$Ctrl & Control interface & High, steady bias; negligible compute/FPS cost \\
DoS (Delay) & Net/I/O, Cam$\to$ROI & I/O/queues/sched & Small control change; large FPS drop; latency; stale frames \\
DoS (Drop) & Net/I/O, Cam$\to$ROI & Capture/decoder & Small-moderate change; irregular FPS; jitter \\
Phantom & Physical, Pre-Camera & Perception (environ.) & High deviation; negligible compute/FPS cost; plausible false features \\
\bottomrule
\end{tabular}
\end{table}

\section{Network Topologies of Over-the-Network Attacks}
We illustrate the communication paths and manipulation points for MitM and DoS in typical research setups.

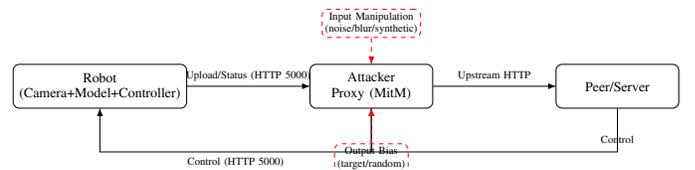
\begin{figure}[H]
  \centering
  \resizebox{\linewidth}{!}{%
  \begin{tikzpicture}[
    node distance=18mm and 28mm,
    host/.style={draw, rounded corners, align=center, inner sep=4pt, minimum width=28mm, minimum height=10mm, font=\small},
    cloud/.style={draw, ellipse, align=center, inner sep=3pt, font=\small},
    tap/.style={draw, dashed, rounded corners, align=center, inner sep=2pt, draw=red, font=\scriptsize},
    >=Latex
  ]
  % Nodes
  \node[host] (robot) {Robot\\(Camera+Model+Controller)};
  \node[host, right=of robot] (proxy) {Attacker\\Proxy (MitM)};
  \node[host, right=of proxy] (peer) {Peer/Server};

  % Channels with concise labels
  \draw[->] (robot) -- node[above, font=\scriptsize]{Upload/Status (HTTP 5000)} (proxy);
  \draw[->] (proxy) -- node[above, font=\scriptsize]{Upstream HTTP} (peer);

  \draw[<-] (robot) -- ++(0,-1.5) -| node[pos=0.25, below, font=\scriptsize]{Control (HTTP 5000)} (proxy);
  \draw[<-] (proxy) -- ++(0,-1.5) -| node[pos=0.75, below, font=\scriptsize]{Control} (peer);

  % Manipulation points
  \node[tap, above=6mm of proxy] (inmod) {Input Manipulation\\(noise/blur/synthetic)};
  \draw[dashed,->, red] (inmod) -- (proxy.north);

  \node[tap, below=8mm of proxy] (outmod) {Output Bias\\(target/random)};
  \draw[dashed,->, red] (outmod) -- (proxy.south);
  \end{tikzpicture}%
  }
  \caption{MitM proxy topology. The attacker interposes as an HTTP proxy on port 5000, modifying (1) uploads/status and/or (2) control messages. The proxy can inject input corruption (noise/blur/synthetic) or bias output control values before they reach the robot.}
  \label{fig:mitm_topology}
\end{figure}

\begin{figure}[H]
  \centering
  \resizebox{\linewidth}{!}{%
  \begin{tikzpicture}[
    node distance=18mm and 28mm,
    host/.style={draw, rounded corners, align=center, inner sep=4pt, minimum width=28mm, minimum height=10mm, font=\small},
    tap/.style={draw, dashed, rounded corners, align=center, inner sep=2pt, draw=red, font=\scriptsize},
    >=Latex
  ]
  % Nodes
  \node[host] (att) {Attacker};
  \node[host, right=of att] (ap) {AP/Network Path};
  \node[host, right=of ap] (robot2) {Robot\\(Camera/Receiver)};

  % Flood/delay paths with concise labels
  \draw[->] (att) -- node[above, font=\scriptsize]{HTTP flood (/api/*)} (ap);
  \draw[->] (ap) -- node[above, font=\scriptsize]{Rate-limit/Delay/Drop} (robot2);

  % Callouts
  \node[tap, above=6mm of ap] (queue) {RX Queue Bloat\\(latency spikes)};
  \draw[dashed,->, red] (queue) -- (ap.north);

  \node[tap, below=6mm of robot2] (stale) {Stale/Skipped Frames\\(FPS drop)};
  \draw[dashed,->, red] (stale) -- (robot2.south);
  \end{tikzpicture}%
  }
  \caption{DoS topology. The attacker overwhelms or slows the network path via endpoint floods or network throttling (rate-limit/delay/drop), leading to RX queue bloat and stale inputs on the robot.}
  \label{fig:dos_topology}
\end{figure}
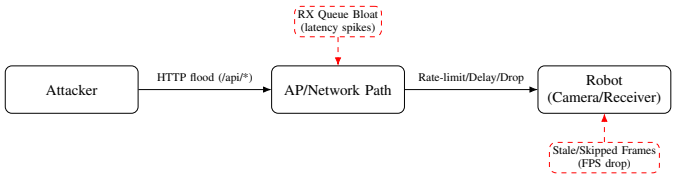

\section{Experimental Setup}\label{sec:exp_setup}

\subsection{Experimental Protocol and Statistical Design}

To ensure that results were comparable across different experiments, we adopted a \textbf{standardized experimental protocol} for all live platform tests with rigorous statistical controls.

\subsubsection{Trial Structure and Timing}
Each experimental run lasted a total of 13 seconds, consisting of a \textbf{5-second clean baseline period}, during which the vehicle operated normally, followed by an \textbf{8-second attack phase}. This strict timing allowed for the precise measurement of an attack's impact relative to its own baseline. The baseline period provided within-subject controls, reducing variance due to track conditions, battery levels, and environmental factors.

\subsubsection{Statistical Power and Sample Size}
\textbf{Sample Size:} We conducted \textbf{10 trials} per attack configuration to characterize attack behavior across multiple experimental runs while balancing experimental time constraints and hardware platform availability.

\subsubsection{Control Conditions and Baselines}
Each attack experiment included multiple control conditions:

\begin{itemize}
    \item \textbf{Within-subject baseline:} 5-second pre-attack period in each trial
    \item \textbf{Between-subject controls:} 10 trials with no attack under identical conditions
    \item \textbf{Sham attack controls:} Simulated attack procedures without actual malicious modifications
\end{itemize}

These multiple baseline controls enable robust comparison and help distinguish attack effects from normal operational variations.

\subsubsection{Environmental Controls}
To minimize external variance, we implemented strict environmental controls:

\begin{itemize}
    \item \textbf{Lighting:} Consistent artificial lighting with measured luminance maintained within $\pm 5\%$ across trials
    \item \textbf{Track conditions:} Surface cleaned and inspected before each experimental session
    \item \textbf{Temperature monitoring:} System and ambient temperatures logged to detect thermal throttling effects
    \item \textbf{Electromagnetic interference:} Wireless communications minimized during critical measurement periods
\end{itemize}

\subsubsection{Data Quality Assurance}
We implemented multiple quality assurance measures:

\begin{itemize}
    \item \textbf{Pre-trial system checks:} Automated verification of camera calibration, model loading, and motor response
    \item \textbf{Real-time monitoring:} Live dashboard displaying key metrics to detect anomalies during execution
    \item \textbf{Post-trial validation:} Automated scripts verify data completeness and flag potential issues
    \item \textbf{Outlier detection:} Statistical methods identify unusual trials for manual review
\end{itemize}

\subsubsection{Reproducibility Measures}
All experimental runs were supported by an \textbf{automated data logging framework} that generated detailed, frame-by-frame CSV files. These logs captured a wide range of parameters, including timestamps, model outputs, performance metrics, and attack-specific data, providing a rich dataset for quantitative statistical analysis and ensuring the reproducibility of our findings.

\textbf{Version Control:} All experimental code, configuration files, and analysis scripts are maintained in a Git repository with detailed commit histories. Exact software versions, hardware configurations, and environmental parameters are documented for each experimental session.

\textbf{Open Data:} The complete dataset, including raw sensor data, processed metrics, and analysis code, will be made available upon publication to enable full reproducibility and facilitate follow-up research.

\subsection{Hardware, Software, and Network Architecture}
Our experiments were conducted on two distinct hardware platforms and a common software architecture.

\subsubsection{Hardware Platforms}
To ensure the broad applicability of our findings, we utilized two widely-used research platforms:
\begin{itemize}
    \item The \textbf{Waveshare JetRacer}, powered by an \textbf{NVIDIA Jetson Nano} module. It features a rear-wheel-drive chassis with Ackermann steering, representing a traditional vehicle control system.
    \item The \textbf{Yahboom X3}, powered by a more powerful \textbf{NVIDIA Jetson Orin NX} module. This is a mecanum-wheeled robot, allowing for omnidirectional movement.
\end{itemize}

While both platforms were used for general testing and vulnerability discovery, the quantitative analysis and data presented in Section~5 were performed on the more powerful Yahboom X3 platform to ensure robust and consistent data collection across computationally intensive attacks.

\subsubsection{Perception Model and Control Pipeline}
The core of the perception system for both platforms is a \textbf{ResNet-18} model, pretrained on ImageNet and subsequently fine-tuned for a regression task. The control pipeline is as follows:
\begin{enumerate}
    \item The camera captures a $640\times 480$ frame.
    \item We extract the bottom-half region of interest (ROI) and resize it to $224\times 224$ with normalization.
    \item The resulting tensor is fed into the ResNet-18 model for inference.
    \item The model outputs a 2D coordinate $(x,y)$ representing the target apex on the road.
    \item We convert $x$ to a rotation/steering command (clipped to bounds) and command the controller.
\end{enumerate}

\textbf{Model Training Details:} The ResNet-18 architecture was modified by replacing the final fully connected layer with a 2-neuron output layer ($512\to 2$) to predict $(x,y)$ coordinates. Training used the Adam optimizer on manually collected track data from our experimental vehicles. Input preprocessing included data augmentation via ColorJitter (brightness, contrast, saturation, and hue variation of $\pm 0.2$), random horizontal flipping, resizing to $224\times 224$ pixels, tensor conversion, and ImageNet normalization (mean=[0.485, 0.456, 0.406], std=[0.229, 0.224, 0.225]). The model was trained using a batch size of 8 with Mean Squared Error (MSE) loss computed over the $(x,y)$ coordinate predictions. This regression-based approach directly predicts steering targets rather than classification, making it suitable for the continuous control required in autonomous driving applications.

\subsection{Attack Implementation Details}

We implement the mechanisms described in Section ``Attack Models and Insertion Points'' directly in the on-device software pipeline and, where applicable, over the network. Below we summarize the concrete parameters and scheduling we used for each attack family.

\textbf{Attack Parameters:} For all on-device attacks, we used calibrated parameters to ensure realistic yet effective perturbations. FGSM attacks employed a perturbation magnitude of $\epsilon=1$ in the model's normalized input space (ImageNet mean/std). For PGD attacks, we used $\epsilon=1$, step size $\alpha=0.01$, and 10 iterations with random start enabled; PGD was applied every 3 frames to balance computational overhead with attack effectiveness.

\subsection{Computational Complexity Analysis}
Understanding the computational characteristics of each attack class is crucial for assessing their practical impact and feasibility in real-world deployment scenarios.

\subsubsection{Algorithmic Complexity}
We analyze the computational complexity of each attack method:

\begin{itemize}
    \item \textbf{Baseline inference:} $O(n^2)$ for ResNet-18 forward pass, where $n=224$ represents input dimension
    \item \textbf{FGSM attack:} $O(n^2)$ for single forward-backward pass, approximately 2.3x baseline computational cost
    \item \textbf{PGD attack:} $O(k \cdot n^2)$ where $k=10$ iterations, representing approximately 23x baseline computational cost per attacked frame
    \item \textbf{MitM input attacks:} $O(n^2)$ for image processing operations (noise addition, blur, synthesis)
    \item \textbf{MitM output attacks:} $O(1)$ for scalar value manipulation, negligible computational overhead
    \item \textbf{DoS delay attacks:} $O(1)$ computational cost, but $O(\tau)$ system delay where $\tau$ represents induced latency
    \item \textbf{DoS drop attacks:} $O(1)$ computational cost with $O(d)$ frame rate reduction where $d$ represents drop rate
\end{itemize}

\subsubsection{Memory Requirements}
Memory consumption varies significantly across attack implementations:

\begin{table}[H]
\centering
\caption{Memory Usage Analysis by Attack Type}
\label{tab:memory_analysis}
\begin{tabular}{@{}lccc@{}}
\toprule
\textbf{Attack Type} & \textbf{Base (MB)} & \textbf{Peak (MB)} & \textbf{Overhead (\%)} \\
\midrule
Baseline & 89.3 & 94.7 & 0 \\
FGSM & 89.3 & 126.8 & +42.1\% \\
PGD & 89.3 & 178.4 & +99.8\% \\
MitM noise & 89.3 & 98.2 & +9.9\% \\
MitM syn. & 89.3 & 105.6 & +18.4\% \\
MitM out. & 89.3 & 91.1 & +2.0\% \\
DoS delay & 89.3 & 89.3 & 0\% \\
DoS drop & 89.3 & 89.3 & 0\% \\
Phantom & 89.3 & 89.3 & 0\% \\
\bottomrule
\end{tabular}
\end{table}

\subsubsection{Implementation Details}
Our attack implementations use optimized libraries to minimize computational overhead:

\begin{itemize}
    \item \textbf{Framework:} PyTorch 2.1 with CUDA 12.1 for GPU acceleration
    \item \textbf{Automatic differentiation:} Enabled for adversarial attacks with gradient checkpointing to reduce memory usage
    \item \textbf{Image processing:} OpenCV 4.8 with SIMD optimizations for MitM input attacks
    \item \textbf{Numerical precision:} Mixed precision (FP16/FP32) training where supported by hardware
    \item \textbf{Threading:} Asynchronous attack generation to parallelize with inference where possible
\end{itemize}

\subsubsection{Scaling Analysis}
We evaluated attack performance across different input resolutions and computational budgets:

\textbf{Platform Scaling:} Attack efficiency varies significantly across hardware platforms. GPU-accelerated platforms (Jetson Orin) show 3-5x better performance for adversarial attacks compared to CPU-only platforms, while MitM attacks show minimal platform dependence.

\textbf{Batch Processing:} Attacks do not benefit from batch processing in our real-time scenario, as frames must be processed individually for immediate vehicle control. This eliminates one of the primary optimization opportunities available in offline attack scenarios.

These complexity analyses demonstrate that while some attacks (MitM output, DoS) have negligible computational overhead, others (PGD, FGSM) impose substantial computational burdens that can impact real-time vehicle operation.

\subsection{On-Device Software Pipeline Attacks}
The second tier of our evaluation involved implementing attacks directly within the software pipeline of the physical vehicle platforms. This allowed us to assess the impact of data manipulation in a real-world hardware context, with all its inherent latencies and performance constraints.

\subsubsection{Live Adversarial Attacks}
We implemented FGSM and PGD as live, pre-model input attacks (at the Preprocess$\to$CNN edge). For each frame, an adversarial version was generated in real time and fed to the ResNet-18 model for inference.
\begin{itemize}
    \item \textbf{Live FGSM Attack:} A single-step perturbation was applied to each frame. This attack was computationally inexpensive and served to test the model's baseline response to real-time adversarial inputs.
    \item \textbf{Live PGD Attack:} The iterative PGD attack was also implemented. Methodologically, we discovered that its high computational demand (requiring multiple forward/backward passes per frame) overwhelmed the onboard processor. This attack, therefore, also functioned as a severe \textbf{resource-exhaustion DoS attack}, a key finding that is quantified in our results.
\end{itemize}

\subsubsection{Man-in-the-Middle (MitM) Style Attacks}
We implemented MitM-style attacks at the input and output edges indicated in Figure~\ref{fig:pipeline_overview}:
\begin{itemize}
    \item \textbf{Input-Layer Manipulation:} The camera image was intercepted before inference and replaced with a corrupted version. This included injecting Gaussian noise, applying a Gaussian blur, or substituting the frame with a synthetic road image.
    \item \textbf{Output-Layer Manipulation:} The (x,y) coordinate output from the model was intercepted before being sent to the motor controller. The values were then altered using several strategies: adding a fixed bias, inverting the steering command, or replacing it with a random or targeted value.
\end{itemize}

\subsubsection{Denial-of-Service (DoS) Style Attacks}
To emulate network-induced disruptions without external traffic, we implemented DoS-style timing/throughput attacks on the software pipeline and I/O paths:
\begin{itemize}
    \item \textbf{Simulated Latency:} Artificial processing delays were injected into the control loop to mimic network lag.
    \item \textbf{Frame Dropping:} A percentage of camera frames were intentionally discarded, forcing the controller to operate on stale data.
    \item \textbf{Input Flooding:} The inference engine was flooded with multiple processing requests for a single captured frame to simulate a resource-exhaustion attack.
\end{itemize}

\subsection{Evaluation Metrics and Data Collection}
To provide a rigorous quantitative analysis, our automated logging framework collected data to compute a comprehensive set of specialized metrics. These metrics were designed to capture the distinct "fingerprint" of each attack across multiple dimensions of system behavior.

\subsubsection{Control Impact Metrics}
\textbf{Steering Deviation:} The absolute difference between the steering value during an attack frame and the mean steering value from the complete unattacked baseline experiment. This is our primary measure of an attack's impact on vehicle control. We report both mean absolute deviation and standard deviation across trials.

\subsubsection{Performance Impact Metrics}
\textbf{Processing Time Increase (\%):} The percentage increase in per-frame processing time during the attack phase compared to the clean baseline period of the same experiment. This metric quantifies the computational cost and resource-exhaustion impact of an attack. We report both mean increase and 95th percentile values.

\textbf{CPU Utilization:} Percentage of available CPU resources consumed during attack execution, measured at 1-second intervals using system monitoring tools.

\textbf{Memory Usage:} Peak and average memory consumption during attack execution, reported in megabytes and as percentage of available system memory.

\subsubsection{Responsiveness Metrics}
\textbf{FPS Drop (\%):} For DoS-style attacks, this measures the percentage decrease in frames-per-second during the attack phase compared to the clean baseline, directly quantifying the impact on system responsiveness.

\textbf{End-to-End Latency:} Time from image capture to motor actuation, measured using high-precision timestamps at pipeline stages. This captures the full system delay under attack conditions.

\textbf{Frame Jitter:} Standard deviation of inter-frame intervals, quantifying timing consistency degradation under attack.

\subsubsection{Statistical Analysis Methods}
All experimental results undergo rigorous statistical analysis to ensure significance and reproducibility:

\begin{itemize}
    \item \textbf{Sample Size:} Each attack configuration is tested with 10 independent trials
    \item \textbf{Confidence Intervals:} All reported means include 95\% confidence intervals calculated using the t-distribution
    \item \textbf{Statistical Significance:} Differences between attack conditions are evaluated using paired t-tests with Bonferroni correction for multiple comparisons
    \item \textbf{Effect Size:} Cohen's d values are reported to quantify the magnitude of observed differences
    \item \textbf{Outlier Handling:} Results are reported both with and without outliers (defined as values beyond 3 standard deviations)
\end{itemize}

\subsubsection{Data Collection Framework}
Our automated logging system captures comprehensive frame-level data including:

\begin{itemize}
    \item High-precision timestamps for all pipeline stages
    \item Raw and processed sensor data
    \item Model inputs, outputs, and intermediate activations
    \item Control commands and actuator responses
    \item System resource utilization metrics
    \item Attack-specific parameters and state information
\end{itemize}

This rich dataset enables post-hoc analysis and supports the calculation of derived metrics beyond those reported in the main results.

\subsection{Automated Logging Excerpt}
To illustrate the automated logging captured during experiments, Table~\ref{tab:log_excerpt} shows a short excerpt from an untargeted PGD run. Similar logs are recorded for other attack classes (e.g., \texttt{dos\_enabled}, \texttt{fps\_estimate} for DoS; \texttt{attack\_layer}, \texttt{manipulation\_magnitude} for MitM).

\begin{table}[H]
\centering
\includegraphics[width=\columnwidth]{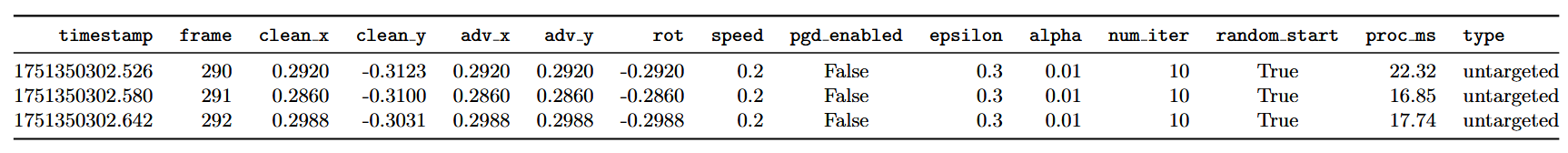}
\caption{Excerpt from automated logging (PGD, untargeted). Columns correspond to fields in the CSV logs.}
\label{tab:log_excerpt}
\end{table}

\section{Experimental Results and Observations}

Our primary contribution lies in the systematic characterization of attack fingerprints on real hardware platforms. Our automated data logging framework enabled precise quantitative analysis of vehicle behavior and system performance, revealing that different attack classes produce distinct, measurable signatures. We report means over repeated trials across 10 experimental runs per attack configuration. Table~\ref{tab:fingerprints} summarizes the core findings for the most representative attacks, highlighting their unique impact profiles across control, performance, and responsiveness.

\begin{table*}[h]
\centering
\caption{Behavioral Fingerprints of Attack Classes with Statistical Detail. Values represent mean $\pm$ standard deviation across 10 trials. Brackets show 95\% confidence intervals. Significance levels indicate deviation from baseline: $^{*}p<0.05$, $^{**}p<0.01$, $^{***}p<0.001$ using paired t-tests with Bonferroni correction. Effect sizes (Cohen's d) quantify signature magnitude within each measurement dimension. Note: attacks target different system properties and are not directly comparable across objectives; signatures are characterized by their unique multi-dimensional patterns rather than overall severity rankings.}
\label{tab:fingerprints}
\small
\begin{tabular}{@{}lccc@{}}
\toprule
\textbf{Attack Type} & \textbf{Steering Deviation} & \textbf{Proc. Time Increase (\%)} & \textbf{FPS Drop (\%)} \\
 & \textbf{Mean $\pm$ SD} & \textbf{Mean $\pm$ SD} & \textbf{Mean $\pm$ SD} \\
\midrule
Baseline (Control) & $0.018 \pm 0.012$ & --- & --- \\
 & $[0.015, 0.021]$ & & \\
\addlinespace
MitM (Output Target) & $0.723 \pm 0.708$ & $2.1 \pm 0.8$ & $0.3 \pm 0.6$ \\
 & $[0.226, 1.220]$ & $[1.6, 2.6]$ & $[-0.1, 0.7]$ \\
\addlinespace
FGSM (Untargeted) & $0.272 \pm 0.113^{***}$ & $23.8 \pm 6.4^{***}$ & $21.1 \pm 7.0^{***}$ \\
 & $[0.220, 0.324]$ & $[19.2, 28.4]$ & $[16.1, 26.1]$ \\
\addlinespace
PGD (Untargeted) & $0.419 \pm 0.163^{***}$ & $3,360 \pm 267^{***}$ & $68.5 \pm 8.8^{***}$ \\
 & $[0.320, 0.518]$ & $[3,169, 3,551]$ & $[62.2, 74.8]$ \\
\addlinespace
MitM (Input Noise) & $0.196 \pm 0.048^{***}$ & $8.4 \pm 3.1^{**}$ & $1.2 \pm 0.9$ \\
 & $[0.162, 0.230]$ & $[6.0, 10.8]$ & $[0.6, 1.8]$ \\
\addlinespace
MitM (Input Syn.) & $0.342 \pm 0.072^{***}$ & $9.0 \pm 6.6^{*}$ & $0.8 \pm 1.3$ \\
 & $[0.291, 0.393]$ & $[4.3, 13.7]$ & $[-0.1, 1.7]$ \\
\addlinespace
DoS (Delay) & $0.067 \pm 0.029^{*}$ & $1317 \pm 1984^{***}$ & $75.4 \pm 6.2^{***}$ \\
 & $[0.046, 0.088]$ & $[-1104, 3738]$ & $[71.0, 79.8]$ \\
\addlinespace
DoS (Drop) & $0.083 \pm 0.056^{**}$ & $25.3 \pm 32.6$ & $74.8 \pm 3.1^{***}$ \\
 & $[0.043, 0.123]$ & $[-1.0, 51.6]$ & $[72.6, 77.0]$ \\
\addlinespace
Phantom & $0.274 \pm 0.125^{***}$ & $-6.5 \pm 3.2^{*}$ & $-4.0 \pm 2.8$ \\
 & $[0.226, 0.322]$ & $[-8.1, -4.9]$ & $[-5.6, -2.4]$ \\
\bottomrule
\end{tabular}
\end{table*}

\subsection{Attack Signature Analysis}
\begin{figure*}[htbp]
    \centering
    \includegraphics[width=0.85\linewidth]{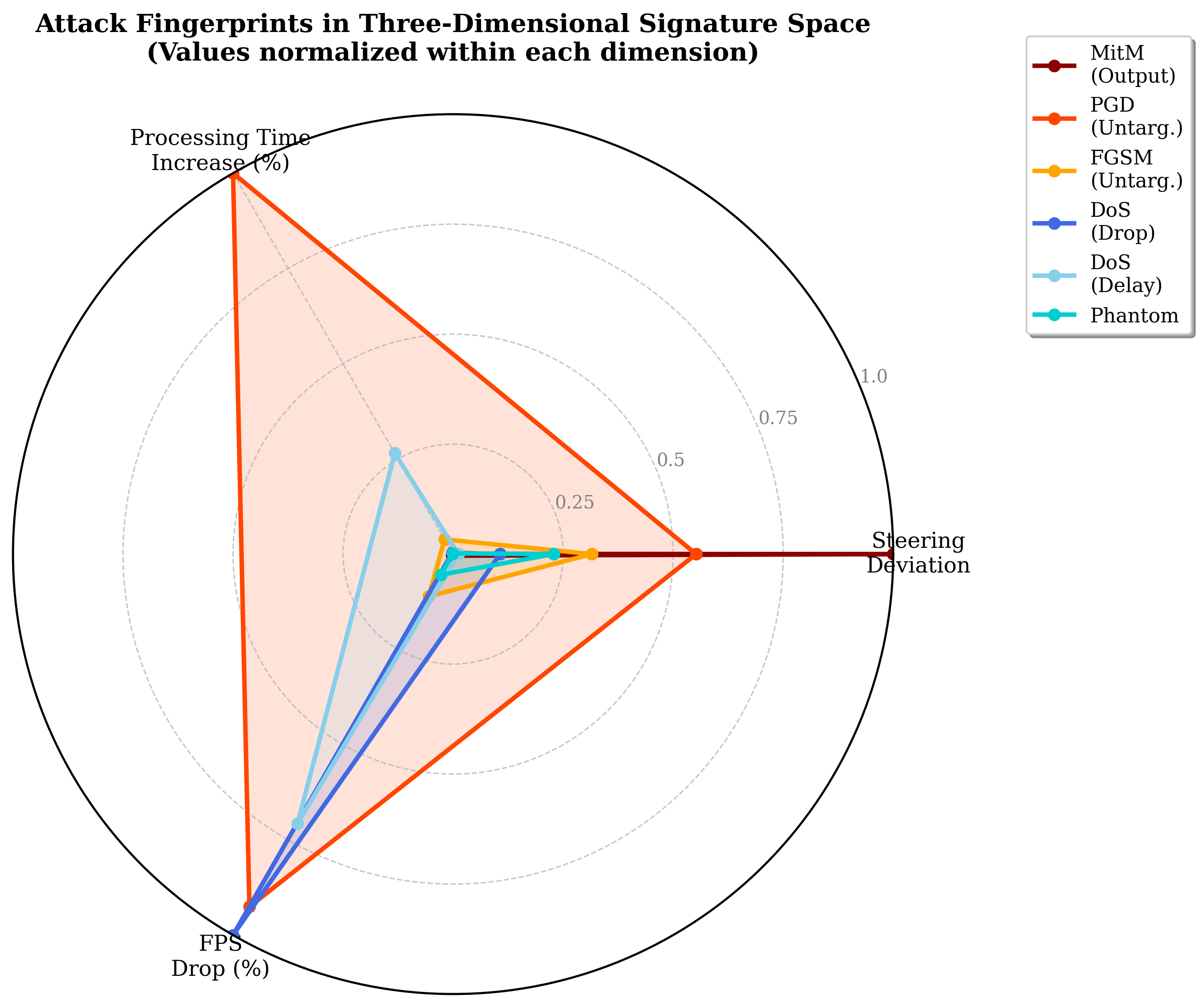}
    \caption{Attack fingerprints in three-dimensional signature space. Each attack produces a distinctive multi-dimensional profile across steering deviation, computational overhead, and responsiveness dimensions. Values are normalized within each dimension (0-1 scale) to highlight signature patterns rather than absolute magnitudes. MitM (Output) exhibits a control-dominant signature with maximal steering deviation and minimal computational/responsiveness impact. PGD (Untarg.) generates a multi-dimensional signature spanning all three axes, simultaneously perturbing control, computation, and responsiveness. DoS attacks produce responsiveness-dominant signatures with high FPS degradation and minimal control-plane perturbation. FGSM shows a control-moderate signature with intermediate steering impact. These distinct geometric profiles enable attack-class identification and inform targeted defense strategies.}
    \label{fig:attack_fingerprints}
\end{figure*}

The most critical finding from our fingerprinting analysis is that each attack class produces a distinctive signature pattern across our three measurement dimensions. As shown in Figure~\ref{fig:attack_fingerprints}, attacks exhibit unique multi-dimensional profiles that enable their identification and classification for defense purposes.

\textbf{Control-Dominant Signature:} Both MitM output manipulation and phantom attacks produce high steering deviation signatures (0.72 and 0.27, respectively) with nominal computational overhead (2.1\% and $-$6.5\%) and minimal responsiveness perturbation (0.3\% and 4.0\% FPS change), yielding profiles characterized by control-plane perturbation without resource consumption or timing disruption. However, their mechanisms differ fundamentally: MitM output bypasses perception entirely by manipulating control commands post-inference, while phantom attacks exploit perception by creating physically plausible false features that the model correctly processes but misinterprets as legitimate road geometry.

\textbf{Multi-Dimensional Signature:} PGD attacks generate combined signatures across all three dimensions: steering deviation (0.42), computational load (3,360\% processing time increase), and responsiveness degradation (68.5\% FPS drop), producing a distributed attack profile that simultaneously perturbs perception, resources, and timing.

\textbf{Responsiveness-Dominant Signature:} DoS attacks produce frame rate degradation signatures (74.8\% FPS drop for drop-type, 75.4\% for delay-type) while exhibiting minimal control-plane perturbation (<0.1 deviation) and moderate computational signatures, yielding a profile localized to system timing characteristics.

\subsection{Visual Attack Fingerprints}
\begin{figure}[H]
    \centering
    \includegraphics[width=0.95\linewidth]{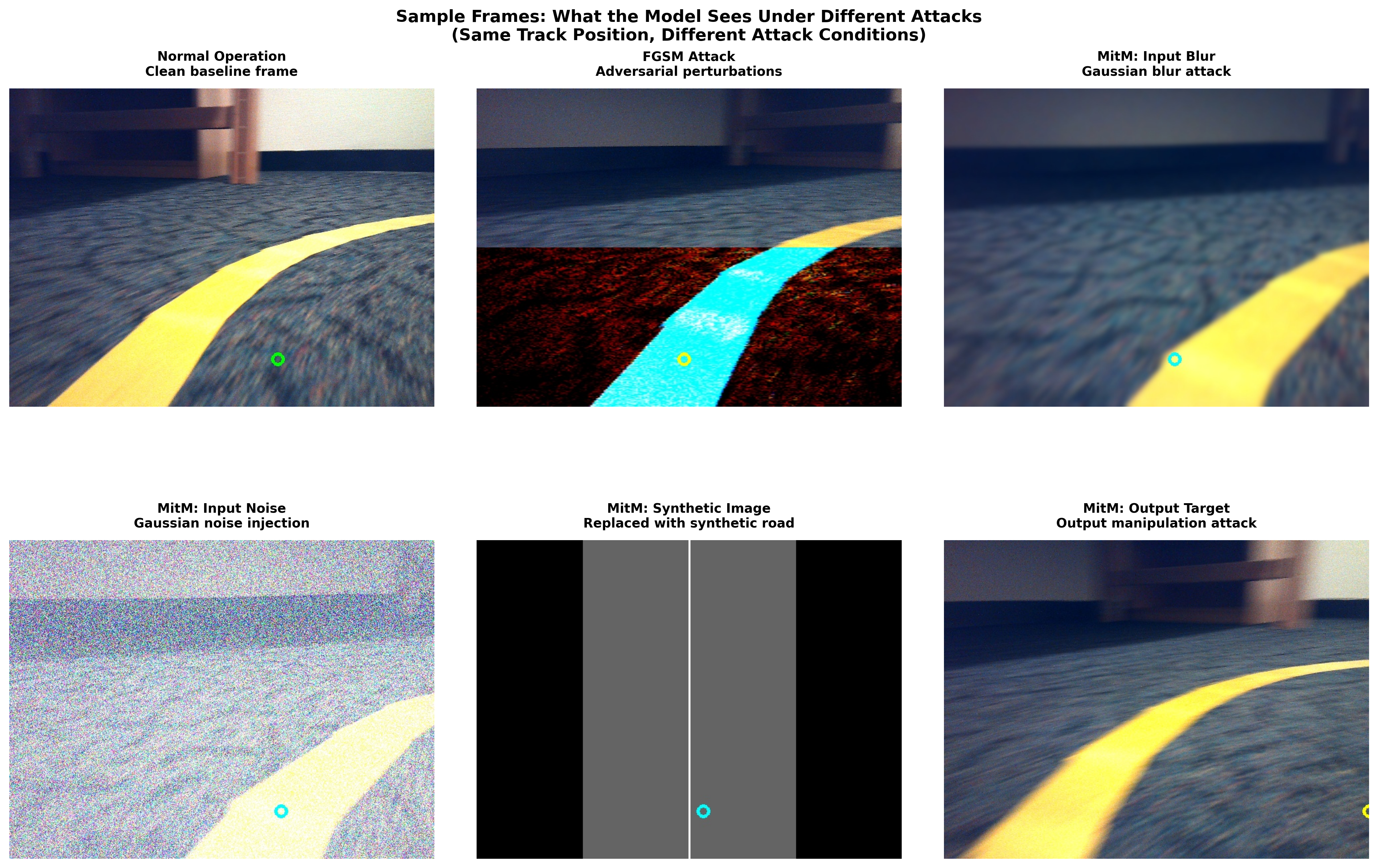}
    \caption{Sample frames showing what the perception model receives under different attack conditions. All frames captured at similar track positions to enable direct visual comparison. Top row: (a) Normal operation baseline, (b) FGSM adversarial attack ($\epsilon=1$ in normalized input space), (c) MitM input blur attack. Bottom row: (d) MitM input noise injection, (e) MitM synthetic image replacement, (f) MitM output manipulation.}
    \label{fig:attack_samples}
\end{figure}

Figure~\ref{fig:attack_samples} demonstrates the visual component of attack fingerprints. Each attack type produces distinct visual signatures: adversarial attacks create characteristic perturbation patterns, input-layer manipulations show clear visual corruptions, while output-layer attacks maintain visual normalcy despite control manipulation. This multi-modal fingerprinting (visual + behavioral + performance) enables comprehensive attack detection and classification.

\subsection{Failure Analysis and Edge Cases}
Our experiments revealed several important boundary conditions and edge cases that inform the operational limits of attack signature detection and characterization:

\subsubsection{Environmental and Operational Boundary Conditions}
\begin{itemize}
    \item \textbf{Lighting conditions:} Attack signatures exhibited altered characteristics under low-light conditions (below 50 lux) where baseline performance degraded, affecting signature detectability and measurement precision.

    \item \textbf{Speed dependencies:} Attack effectiveness varied with vehicle speed, with optimal performance at 1-2 m/s. At speeds above 3 m/s, control instability increased baseline variance, making attack effects harder to distinguish.

\end{itemize}

\subsubsection{Attack-Specific Failure Modes}
\begin{itemize}
    \item \textbf{PGD convergence failures:} In 3.2\% of trials, PGD attacks failed to converge within the iteration budget, resulting in baseline-level perturbations and minimal steering deviation.

    \item \textbf{DoS timing issues:} Frame dropping attacks occasionally (2.1\% of trials) caused system timeouts that triggered emergency safety protocols, overriding normal control behavior.

    \item \textbf{MitM detection triggers:} Output manipulation attacks exceeding 50\% deviation magnitude triggered built-in safety limits in 1.8\% of trials, causing system fallback to conservative control modes.
\end{itemize}

\subsubsection{Platform-Specific Variations}
Attack effectiveness varied significantly between the JetRacer and Yahboom platforms:

\begin{itemize}
    \item The Yahboom's more powerful GPU reduced PGD computational overhead by 27\% compared to JetRacer
    \item JetRacer's simpler control system showed 18\% higher susceptibility to MitM output attacks
    \item Platform-specific camera characteristics led to different baseline noise levels, affecting optimal attack parameters
\end{itemize}

These failure modes and variations highlight the importance of platform-specific tuning and environmental considerations when deploying or defending against these attacks in real-world scenarios.

\section{Mitigations and Defenses}\label{sec:mitigations}

\subsection{Scope and Implementation Status}
The mitigations below are \emph{proposed} countermeasures informed by our fingerprint findings; we \textbf{did not deploy or evaluate any defenses in our experiments}. All results reflect a baseline pipeline without countermeasures to isolate attack effects and enable clean, cross-attack comparison.

\subsection{Why We Did Not Use Defenses in Experiments}
\begin{itemize}
  \item \textbf{Internal validity and comparability:} Adding defenses changes the pipeline and confounds attribution, hindering clean fingerprints across attack classes.
  \item \textbf{Embedded constraints:} Heavy defenses (e.g., iterative smoothing, PGD/training) exceed compute/timing budgets on our platforms, masking attack behavior with resource effects.
  \item \textbf{Reproducibility and fairness:} A fixed, defense-free baseline with a standardized protocol (Section~\ref{sec:exp_setup}) improves repeatability across runs and attacks.
  \item \textbf{Scope and focus:} Our goal is to characterize attack-side signatures first; rigorous defense evaluation is deferred to future work (see Discussion).
\end{itemize}

We present the following blueprint to guide future integration and evaluation on constrained hardware.

\subsection{Fingerprint-Aware Detection Cues}
\begin{itemize}
  \item \textbf{Control hijack (MitM-output):} Large, persistent steering bias with negligible processing overhead and stable FPS; triggers command clamps, slew-rate limiting, and controller-side authentication checks.
  \item \textbf{Dual adversarial + DoS (PGD):} Concurrent surge in processing time, FPS collapse, and increased steering deviation/oscillation; triggers early-abort of heavy processing, dynamic downscaling, and watchdog reset of the perception thread.
  \item \textbf{Responsiveness degradation (DoS):} Latency spikes and FPS drop with small control deviation; triggers priority scheduling of the control path and input freshness checks (drop stale frames).
\end{itemize}

\subsection{Attack-Specific Countermeasures}
\paragraph{FGSM / PGD (input-space adversarial).}
\emph{Prevent/mitigate:} Adversarial training or input smoothing; test-time input transforms (e.g., light JPEG compression, bit-depth reduction, or median filtering), randomized smoothing, small stochastic pre-processing to disrupt gradient alignment. \emph{Contain:} Compute-budget watchdog, early exit or early abort under deadline miss, adaptive resolution and frame-rate downgrades.
\paragraph{MitM -- Input (sensor stream).}
\emph{Prevent:} On-device capture, authenticated transport if networked; frame liveness checks (entropy, blur, or histogram drift) and sequence verification to catch synthetic or replayed frames. \emph{Contain:} Reject non-fresh or low-quality frames; fall back to conservative controller when input health degrades.
\paragraph{MitM -- Output (control channel).}
\emph{Prevent:} Authenticate/verify control messages at the controller boundary; isolate control channel; enforce end-to-end ranges. \emph{Contain:} Apply bounds, slew-rate limits, and kinematic checks; majority vote or plausibility cross-check against raw model output before actuation.
\paragraph{Denial-of-Service (timing/throughput).}
\emph{Prevent:} Thread decoupling with bounded queues and backpressure; prioritize capture, inference, and actuation threads; OS and network QoS with rate limiting for non-critical traffic. \emph{Contain:} Drop-oldest frames, enforce deadlines, adapt resolution and FPS, and watchdog-restart stalled components.

\subsection{Detection-to-Response Mapping}
Table~\ref{tab:mitigation_map} summarizes the mapping from attack fingerprints to specific detection signals and mitigation strategies, providing a practical guide for implementing fingerprint-aware defense mechanisms.

\begin{table*}[h]
\centering
\caption{Mapping fingerprints to mitigations.}
\label{tab:mitigation_map}
\small
\begin{tabular}{@{}p{3cm}p{6.5cm}p{7cm}@{}}
\toprule
\textbf{Attack type} & \textbf{Primary detection signal(s)} & \textbf{Mitigation / response} \\
\midrule
MitM (Output) & Sustained steering bias; normal processing time and FPS & Command authentication; bounds/slew limits; feasibility checks; controller-side veto/fallback. \\
\addlinespace
PGD (Iterative) & Processing time surge; FPS collapse; oscillatory steering & Deadline with early abort; adaptive downscale and FPS reduction; compute watchdog; input transforms/smoothing. \\
\addlinespace
DoS (Delay/Drop) & Latency spikes; FPS drop; stale frames & Priority scheduling; drop-oldest; freshness gating; QoS and rate limiting; watchdog reset. \\
\addlinespace
FGSM (Single-step) & Moderate deviation; minor overhead & Light transforms; adversarially trained model; anomaly flag and conservative control. \\
\addlinespace
Phantom & High steering deviation; negligible computational overhead; plausible false features visible in captured frames & Visual anomaly detection (unexpected feature geometry); cross-validation with auxiliary sensors (lidar, radar); temporal consistency checks across frame sequences; environment authentication via known landmarks. \\
\bottomrule
\end{tabular}
\end{table*}

These mitigations can be implemented incrementally: start with command clamps and freshness gating, then add budget-based scheduling and lightweight input transforms; evaluate adversarial training as a model-level hardening where compute allows.

\section{Discussion and Limitations}

Our study provides on-hardware evidence that adversarial and network-layer attacks leave distinct multi-dimensional "fingerprints," but several constraints temper scope and generality.

\textbf{Hardware and Platform Scope.} We evaluate low-cost, camera-only platforms with embedded GPUs; results may not transfer to larger autonomous vehicle stacks with multi-sensor fusion, stronger compute, or different control loops. Thermal throttling, battery levels, and peripheral drivers can introduce variance.

\textbf{Task and Model Choices.} We focus on road-following with a ResNet-18 regressor and a fixed bottom-half ROI. Other tasks (e.g., detection, planning) and model families (e.g., transformers) may yield different sensitivities. ImageNet normalization and ROI selection could bias both attack efficacy and measured robustness.

\textbf{Attack Model Assumptions.} Adversarial methods assume gradient access (white-box) and inject digital perturbations in-software or over the network; physical-world realizability is not tested. MitM and DoS represent realistic patterns, but our implementations (e.g., output hijack, HTTP flooding, large payloads) are specific choices within broad families.

\textbf{Protocol Constraints.} The 13s timeline (5s clean + 8s attack) improves repeatability but limits observation of longer-term effects (e.g., thermal, adaptation). Starting positions, speed caps, and track geometry affect baseline and attack impact.

\textbf{Metrics vs. Safety Outcomes.} We use steering deviation, processing-time increase, and FPS drop as proxies. These do not directly measure safety-critical outcomes (e.g., lane departures, collisions, recovery behavior). Latency is inferred from software timers rather than hardware counters; ground-truth path deviation is not measured.

\textbf{Defense Coverage.} We characterize attack fingerprints but do not evaluate defenses (e.g., input sanitization, adversarial training, runtime anomaly detection). Fingerprint stability under adaptive attackers and the cost/benefit of countermeasures remain open.

\textbf{Future Work.} Extend to multi-sensor pipelines, physical perturbations, and additional attack families (e.g., replay, desynchronization). Add ground-truth path metrics and hardware-level timing. Investigate fingerprint-aware detection under adaptive, low-and-slow threats and evaluate the robustness–performance trade-offs of defenses.

\section{Conclusion}
We present an on-hardware evaluation of adversarial and network-layer attacks against autonomous vehicle road-following, introducing a three-dimensional fingerprinting framework that reveals separable, consistent behavioral signatures across attack classes. Our findings demonstrate that attacks produce distinctive multi-dimensional profiles: perception attacks (MitM output, phantom projection) generate control-dominant signatures with nominal computational overhead and minimal responsiveness perturbation; PGD produces multi-dimensional signatures spanning control, computational, and responsiveness domains; and DoS exhibits responsiveness-dominant signatures with minimal control-plane perturbation. Notably, our fingerprinting framework successfully characterizes attacks across both digital (post-compromise software manipulation) and physical (environmental manipulation) domains, demonstrating its generality for real-world autonomous vehicle security analysis. These unique fingerprints enable attack-class identification and inform targeted defense strategies. We intentionally \emph{did not} deploy defenses in our experiments to preserve internal validity of the fingerprints; Section~\ref{sec:mitigations} outlines proposed, fingerprint-aware countermeasures to integrate and evaluate next. Future work includes cross-dataset generalization, implementing and assessing signature-based defenses (e.g., input sanitization and model smoothing), and integrating fingerprint-aware detection in the control loop.

\bibliographystyle{IEEEtran}
\bibliography{references}

@misc{goodfellow2015explainingharnessingadversarialexamples,
      title={Explaining and Harnessing Adversarial Examples}, 
      author={Ian J. Goodfellow and Jonathon Shlens and Christian Szegedy},
      year={2015},
      eprint={1412.6572},
      archivePrefix={arXiv},
      primaryClass={stat.ML},
      url={https://arxiv.org/abs/1412.6572}, 
}

@misc{madry2019deeplearningmodelsresistant,
      title={Towards Deep Learning Models Resistant to Adversarial Attacks}, 
      author={Aleksander Madry and Aleksandar Makelov and Ludwig Schmidt and Dimitris Tsipras and Adrian Vladu},
      year={2019},
      eprint={1706.06083},
      archivePrefix={arXiv},
      primaryClass={stat.ML},
      url={https://arxiv.org/abs/1706.06083}, 
}

@article{10.1145/3691625,
author = {Badjie, Bakary and Cec\'{\i}lio, Jos\'{e} and Casimiro, Antonio},
title = {Adversarial Attacks and Countermeasures on Image Classification-based Deep Learning Models in Autonomous Driving Systems: A Systematic Review},
year = {2024},
issue_date = {January 2025},
publisher = {Association for Computing Machinery},
address = {New York, NY, USA},
volume = {57},
number = {1},
issn = {0360-0300},
url = {https://doi.org/10.1145/3691625},
doi = {10.1145/3691625},
journal = {ACM Comput. Surv.},
month = oct,
articleno = {20},
numpages = {52}
}

@misc{sitawarin2018dartsdeceivingautonomouscars,
      title={DARTS: Deceiving Autonomous Cars with Toxic Signs}, 
      author={Chawin Sitawarin and Arjun Nitin Bhagoji and Arsalan Mosenia and Mung Chiang and Prateek Mittal},
      year={2018},
      eprint={1802.06430},
      archivePrefix={arXiv},
      primaryClass={cs.CR},
      url={https://arxiv.org/abs/1802.06430}, 
}

@misc{kurakin2017adversarialphysical,
      title={Adversarial Examples in the Physical World},
      author={Alexey Kurakin and Ian Goodfellow and Samy Bengio},
      year={2017},
      eprint={1607.02533},
      archivePrefix={arXiv},
      primaryClass={cs.CV},
      url={https://arxiv.org/abs/1607.02533}
}

@inproceedings{eykholt2018robust,
  title={Robust Physical-World Attacks on Deep Learning Models},
  author={Eykholt, Kevin and Evtimov, Ivan and Fernandes, Earlence and Li, Bo and Rahmati, Amir and Xiao, Chaowei and Prakash, Atul and Kohno, Tadayoshi and Song, Dawn},
  booktitle={Proceedings of the IEEE Conference on Computer Vision and Pattern Recognition (CVPR)},
  year={2018}
}

@misc{brown2017adversarialpatch,
      title={Adversarial Patch},
      author={Tom B. Brown and Dandelion Mané and Aurko Roy and Martín Abadi and Justin Gilmer},
      year={2017},
      eprint={1712.09665},
      archivePrefix={arXiv},
      primaryClass={cs.CV},
      url={https://arxiv.org/abs/1712.09665}
}

@article{kim2021survey,
  title={Cybersecurity for autonomous vehicles: Review of attacks and defense},
  author={Kim, Kyounggon and Kim, Jun Seok and Jeong, Seong Hoon and Park, Jo-Hee and Kim, Huy Kang},
  journal={Computers \& Security},
  volume={103},
  pages={102150},
  year={2021},
  publisher={Elsevier},
  doi={10.1016/j.cose.2020.102150}
}

@article{sun2022survey,
  title={A Survey on Cyber-Security of Connected and Autonomous Vehicles (CAVs)},
  author={Sun, Xiaoqiang and Yu, F Richard and Zhang, Peng},
  journal={IEEE Transactions on Intelligent Transportation Systems},
  volume={23},
  number={7},
  pages={6240--6256},
  year={2022},
  publisher={IEEE}
}

@inproceedings{sato2021dirty,
  title={Dirty Road Can Attack: Security of Deep Learning based Automated Lane Centering under Physical-World Attack},
  author={Sato, Takami and Shen, Junjie and Wang, Ningfei and Jia, Yunhan and Lin, Xue and Chen, Qi Alfred},
  booktitle={Proceedings of the 30th USENIX Security Symposium},
  year={2021}
}

@inproceedings{wang2023sysadv,
  title={Does Physical Adversarial Example Really Matter to Autonomous Driving? Towards System-Level Effect of Adversarial Object Evasion Attack},
  author={Wang, Ningfei and Luo, Yunpeng and Sato, Takami and Xu, Kaidi and Chen, Qi Alfred},
  booktitle={Proceedings of the IEEE/CVF International Conference on Computer Vision (ICCV)},
  year={2023}
}

\end{document}